\pdfoutput=1
\documentclass{svjour3}                    % onecolumn
\usepackage{graphicx}
\usepackage{epstopdf}
\begin{document}

\title{Observation of a Dislocation Related Interfacial Friction Mechanism in Mobile Solid $^4$He
}
\thanks{This work was supported by The Israel Science Foundation, grant 1089/13 and by the Technion Fund for Research }


\titlerunning{Interfacial Friction in Mobile Solid $^4$He}        % if too long for running head

\author{Anna Eyal        \and
        Ethan Livne			\and
		Emil Polturak		%etc.
}


\institute{Anna Eyal \at
              Department of Physics, Technion - Israel Institute of Technology, Haifa 32000, Israel\\
			  %\email{livne@physics.technion.ac.il}
			\and
              Ethan Livne \at
              Department of Physics, Technion - Israel Institute of Technology, Haifa 32000, Israel\\
              \and
			 Emil Polturak \at
			Department of Physics, Technion - Israel Institute of Technology, Haifa 32000, Israel
              Tel.: +972-8-8292761\\
              Fax: +972-8-8292027\\
			\\\email{emilp@physics.technion.ac.il}
}

\date{Received: date / Accepted: date}

\maketitle

\begin{abstract}
We report a study of the temperature and stress
dependence of the friction associated with a relative motion of two masses of solid $^4$He in contact.
The situation where "two masses" coupled only by friction exists emerges spontaneously during a disordering of a single crystal contained inside a
annular sample space of torsional oscillator (TO). Under the torque applied by the oscillating walls of the the TO
these "masses" move relative to each other, generating measurable dissipation at their interface. We studied this dissipation between 0.5K and 1.8K in solid samples grown from commercially pure $^4$He and from a 100 ppm $^3$He-$^4$He mixture. The data were analyzed by
modelling the TO as a driven harmonic oscillator. In this model, analysis of the resonant frequency and amplitude of the TO yields the temperature dependence of the friction coefficient. By fitting the data to specific
forms, we found that over our temperature range, the dominant friction mechanism associated with the interfacial motion
results from climb of individual dislocations. The characteristic energy scale
associated with this internal friction is between 3K and 6K, depending on the sample. The fact that a single value of this energy accounts for the data of a given sample supports the idea that the interface between the moving and static solid is well oriented. The relative motion of the solid in this case can perhaps be described as the low limit of "slip-stick" motion.

\end{abstract}

\section{Introduction}
\label{intro}

Investigation of the unusual properties of solid $^4$He as a quantum solid has been a subject of
intense study in recent years. The ongoing search for supersolidity
at low temperatures\cite{a001} which is currently under debate \cite{a002,a003,a004},
stimulated ongoing effort to explore and understand the unique elastic \cite{a005} and
plastic \cite{a006,a007,a008} properties of crystalline He
at very low temperatures. Our group is a using a torsional oscillator, as in the search for supersolidity, however we work at higher temperatures, between 0.5K and 2K, so that our work is not directly related to
this issue which has been searched for at temperature of $\sim$0.2K and lower.
We are particularly interested in understanding the friction mechanism between
crystallites of He moving past each other. The reason why this problem may be interesting is
that a priori, it is not obvious that friction between quantum solids would be the same as
between classical solids. For example, there are additional predictions regarding what are the friction mechanisms in the quantum regime. One such proposed quantum friction
mechanism comes from the Van der Waals force, due to zero point charge
fluctuations \cite{a009}. Another dissipation mechanism involves phonons,
where the phonons populating one of the masses are Doppler
shifted due to the relative motion \cite{a010}. Internal friction in a static
polycrystalline solid He was investigated in the past, and was
found to result from dynamics of pinned dislocations as described by
the Lucke-Granato model \cite{a011,a012}. However, the classical
friction problem of one solid mass moving against another, was not yet
addressed in solid He. Since solid He exists
only under pressure, the classical layout of a friction
experiment with two masses on a tabletop cannot be realized. By chance, we have we found a way to perform such experiments.

\section{Experimental Background}
\label{Experimental}
\begin{figure}[]
\includegraphics[width=3in]{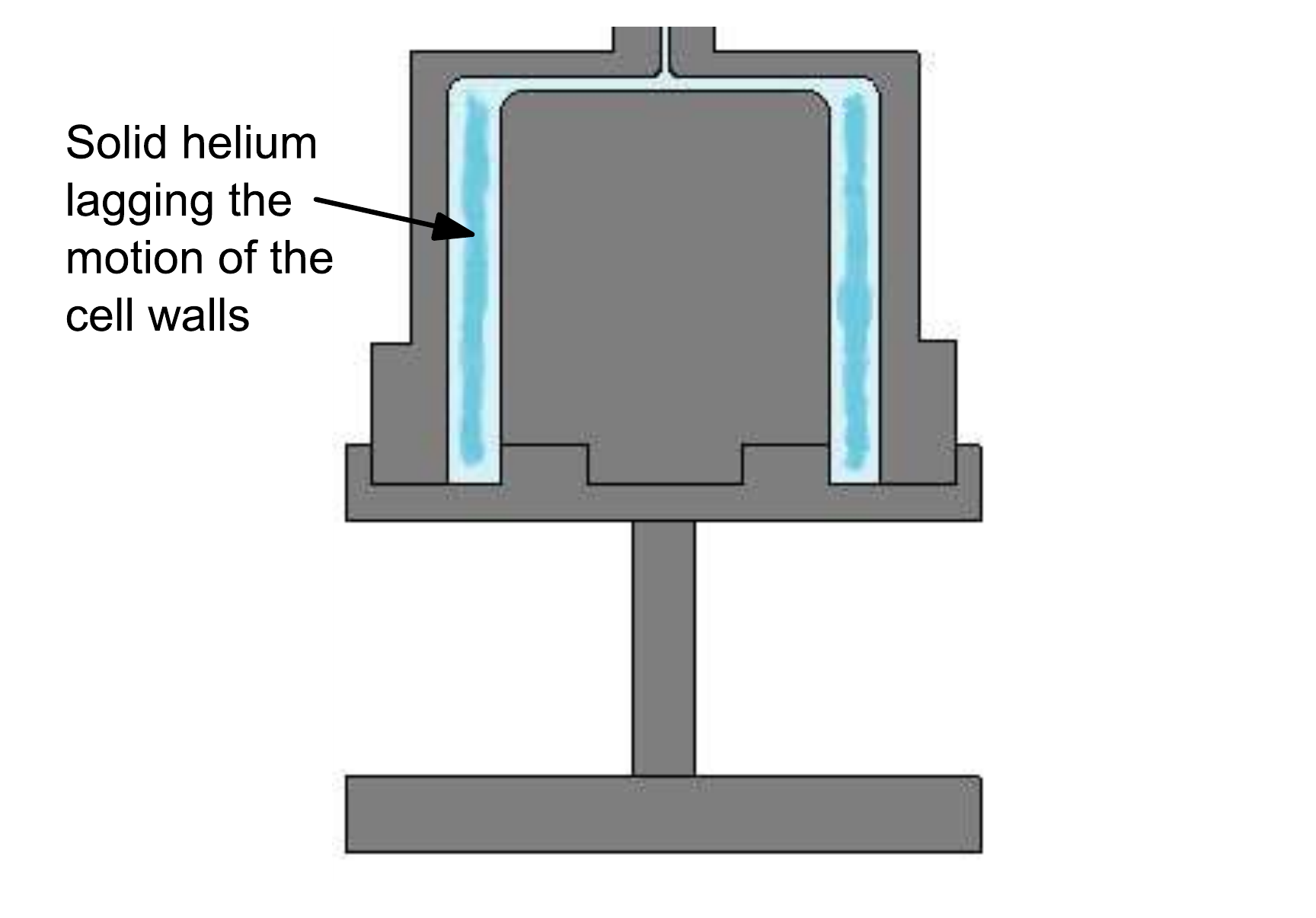} {}
\caption{A schematic cross section of the TO with the annular
sample space filled with disordered solid. The outline of the TO is drawn to scale(see text for dimensions). The outer and inner
walls of the annulus oscillate together. In our model, we divide
the solid inside the annulus into 2 concentric bands; one band contains the solid which is strongly attached
to the outer and inner walls. This part of the solid moves with
the walls. The second band, marked by the arrow, is coupled
to the the solid moving with the walls only by friction. This band
lags the motion of the walls. } \label{fig:grains}
\end{figure}

Our understanding of the situation in which two bodies of solid He move past each other was developed over several years of systematic experiments done in our lab. These experiments are described in great detail in previous publications of our group\cite{a013,a014,a015,a016}. Here, we highlight the main points. Our torsional oscillator is illustrated schematically in Fig.{\ref{fig:grains}}. Solid He is contained inside the annular sample space. Initially, we grow a single crystal of solid He which fills this sample space. Once the sample space is filled with solid, we block the filling line and cool the TO by 50 mK -100 mK. Once cooling begins, the resonant period of the TO decreases spontaneously, as if part of the solid He became decoupled from the motion of the TO. In different experiments, this decoupled part represents between 10\% and 30\% of the moment of inertia of the solid He inside the TO.

The natural question to ask is whether this effect indeed represents real decoupling, where one part of the solid is rigidly attached to the body of the TO and moves with it while another part of the solid is decoupled from the motion of the TO.  This question arises as there could be other reasons responsible for the decrease of the resonant period. Possible "other" reasons include: (a)an appearance of some liquid in the cell, or (b) changes of the elastic properties of the solid with temperature. Regarding (a), when we start to cool the cell, the pressure as measured inside the cell indicates that the TO is completely filled
with solid. Cooling, if anything, converts liquid into solid, not the other way round, so there is no reason why liquid should suddenly appear. To check directly what would be the effect of a residual liquid, we intentionally introduced some liquid into the cell containing solid in the decoupled state. Once the liquid went into the cell, the period of the TO returned to that of a cell filled with a single crystal, and the decoupling effect simply
vanished. So, the mass decoupling cannot be explained by a presence of liquid. By the way, this "residual liquid" test occurs naturally when we grow bcc crystals. As we cool the solid filled cell, its coordinates on the phase diagram move along the hcp-bcc coexistence line and eventually reach the triple point at T=1.446K. At the triple point, liquid appears spontaneously in the cell and immediately, the decoupling effect vanishes. Once we cool below the triple point the liquid disappears and the decoupling reappears. Such observations are shown in Fig.4 of ref. 13. To conclude, the presence of liquid is inconsistent with the observed decrease of the resonant period of the TO.

We now address option (b), namely that the decrease of the resonant period is due to
change of the effective torsion constant of the TO. A decrease of the resonant period will
take place if the shear modulus of solid He increases. The influence of the elastic properties of solid He
on the resonant period of a TO was discussed by several authors\cite{a017,a018}, in conjunction with the search for supersolidity\cite{a001}
at low temperatures. To see whether this possibility applies in our case,
we carried out a similar analysis of our system and for our temperature range using Finite Element Analysis.
This analysis is described in detail in the Appendix of Ref.16. The result is that various effects related to the shear modulus of
solid He affect the resonant period of the TO by 1 to 2 orders of magnitude less than the changes we observe during the decoupling.
Consequently, effects related to the shear modulus cannot explain our results either. We are
therefore left with conclusion that the effective moment of inertia
associated with a polycrystal is smaller than that of a single crystal.
This will take place if part of the solid He does not move as a rigid body
with the TO and becomes effectively decoupled from the motion of the
walls. In the frame of reference of the TO, this decoupled solid moves
relatively to the body of the TO.

To confirm the existence
 of such relative motion directly, we developed an in-situ acoustic AFM-like sensor\cite{a016}. Our kinematical working hypothesis was that if there is a relative motion of two solid masses along the grain boundary which separates them, the atomic corrugation at the interface would generate vibrations at a frequency $f=V/d$, where $V$ is the relative speed and $d$ is the interatomic spacing. Such vibrations were indeed detected, confirming the existence of relative motion. This experiment further established that the orientation of the interface between the static and moving solid is the (0001) plane of the hcp structure. We remark that in contrast to the kinematics which we understand,the physical mechanism describing the relative motion at the interface separating the two masses is not yet understood. However, from the kinematical point of view it is clear that such motion exists, which is sufficient for the purpose of this paper. In order to extract quantitative information, we need the geometry of the two solids, the coupled and the decoupled, inside the TO. In order to have two solids capable of a relative motion inside an annular channel, the interface between them must be invariant under rotation around the symmetry axis of the TO, otherwise relative motion would not be possible. This requirement is satisfied if the solid is divided into concentric annular "bands", illustrated schematically in Fig.{\ref{fig:grains}}.
%According to PIMC simulations, the interface of most grain boundaries in hcp $^4$He is fluid, 2-3 monolayers thick. In our system, such intergranular fluid would appear spontaneously on the interfaces of the grain boundaries between the. A fluid layer would act as a natural lubricant and may well be responsible for the fact the solid bands can move relative to one another with very low dissipation, something which does not happen with usual solids.

 The magnitude of the friction force between these solid "bands" can be roughly estimated from the measured dissipation of our TO, which is proportional to changes of the quality factor.  The moment of inertia of our TO is about40 gm$\cdot$ cm${^2}$ and the unloaded Q$\sim 5\times 10^5$. At a typical tangential speed of 10 $\mu$m/sec, the stored energy is in the 10$^{-5}$ erg range. The dissipation is temperature dependent.  At temperatures above 1K, the dissipation is in the 10$^{-7}$ erg/sec range. At our lowest temperature of 0.5K, it decreases to 10$^{-9}$ erg/sec, which within our precision is the same as that of an empty cell.
 In the geometry shown in Fig.1, the dissipation due to friction can be approximated as $\langle\int{{\bf{F}}\cdot{\bf{v}}ds}\rangle$, where s is the area of the (annular) interface between the moving and static solid, $\bf{F}$ is the friction force per unit area,$\bf{v}$ is the relative speed and the brackets denote time average.  Using experimental values above 1K and a a typical wall speed $v=10\mu$m/sec, we find $F = 6\times10^{-5}$Pa.  This value is 4 orders of magnitude smaller than the critical shear stress of solid He\cite{a005,a019}. The sensitivity is a direct consequence of the high Q of the TO. To compare, the lowest friction force measured to date is 5$\times$10$^4$ Pa, acting between two incommensurate graphite surfaces at room temperature\cite{a020}. Evidently, the friction force associated with solid He is orders of magnitude smaller, something which may allow us to look for weak mechanisms of friction.

Over the last few years, we studied many tens of solid samples prepared as single crystals under a variety of growth conditions\cite{a013,a014,a015}. Depending on the temperature and pressure at which a crystal is grown, we could grow either bcc or hcp solid. Most crystals were grown using commercially pure $^4$He, and some were grown using a 100 ppm $^3$He-$^4$He mixture. While studying the response of the TO containing the solid, we noticed that the variation of the resonant frequency and dissipation of the TO with temperature and applied drive were similar for all of these samples. This observation indicates that the mechanism responsible
for the dissipation is common for all these solid samples. To try and understand this mechanism,  in
this study we collected and analyzed such data of 5 hcp crystals
grown under a variety of conditions: (a) grown in the hcp phase
from a normal fluid at some temperature above the upper triple
point of 1.772K. (b) grown from the superfluid in the hcp phase
below the lower triple point of 1.464K. (c) grown in the bcc phase
at 1.65K, and then transformed into the hcp structure after having
been cooled below the triple point at 1.46K. (d) same as (a) or
(c), but grown from a $^3$He-$^4$He mixture containing 100 ppm of
$^3$He. The selection of a
diverse variety of initial conditions for the crystal growth was
intended to look for quantitatively similar features in their
response to temperature and drive changes, independent of the
growth conditions. For the crystals and TO
used in this experiment, the decoupling process caused the moment
of inertia of the solid to decrease by typically 30\%.  We remark
in this context that when we grow a polycrystalline solid to begin
with rather than a single crystal\cite{a015}, the
decoupled mass decreased to $\sim$1\%, similar to what
is measured by other groups who study polycrystals grown with the blocked capillary technique \cite{a021,a022,a023,a024}. Hence, starting with a single crystal is essential in maximizing the decoupled mass.

\section{Results and Discussion}
\label{Results}
We use an electrostatically driven TO. The inner radius of the annular sample space is 6.5 mm,
the height is 10 mm, and its width is 2 mm (for additional details
see \cite{a013,a016}). The applied torque $\tau_{0}$ is related to the AC driving voltage
$V_{0}$ through $\tau_{0}=V_{DC} r_{0} C V_{0} /d $, where
$V_{DC}=200V$ is a constant bias on one of the capacitor plates;
$r_{0}=8.5mm$ is the radius of the outer wall of the cell;
$C=2pF$; and $d=400 \mu m$ is the distance between the capacitor
plates. With a $V_{0}=$1mV rms, the torque is $10^{-11}
Nm$.

\begin{figure}[]
\includegraphics[width=3.5in]{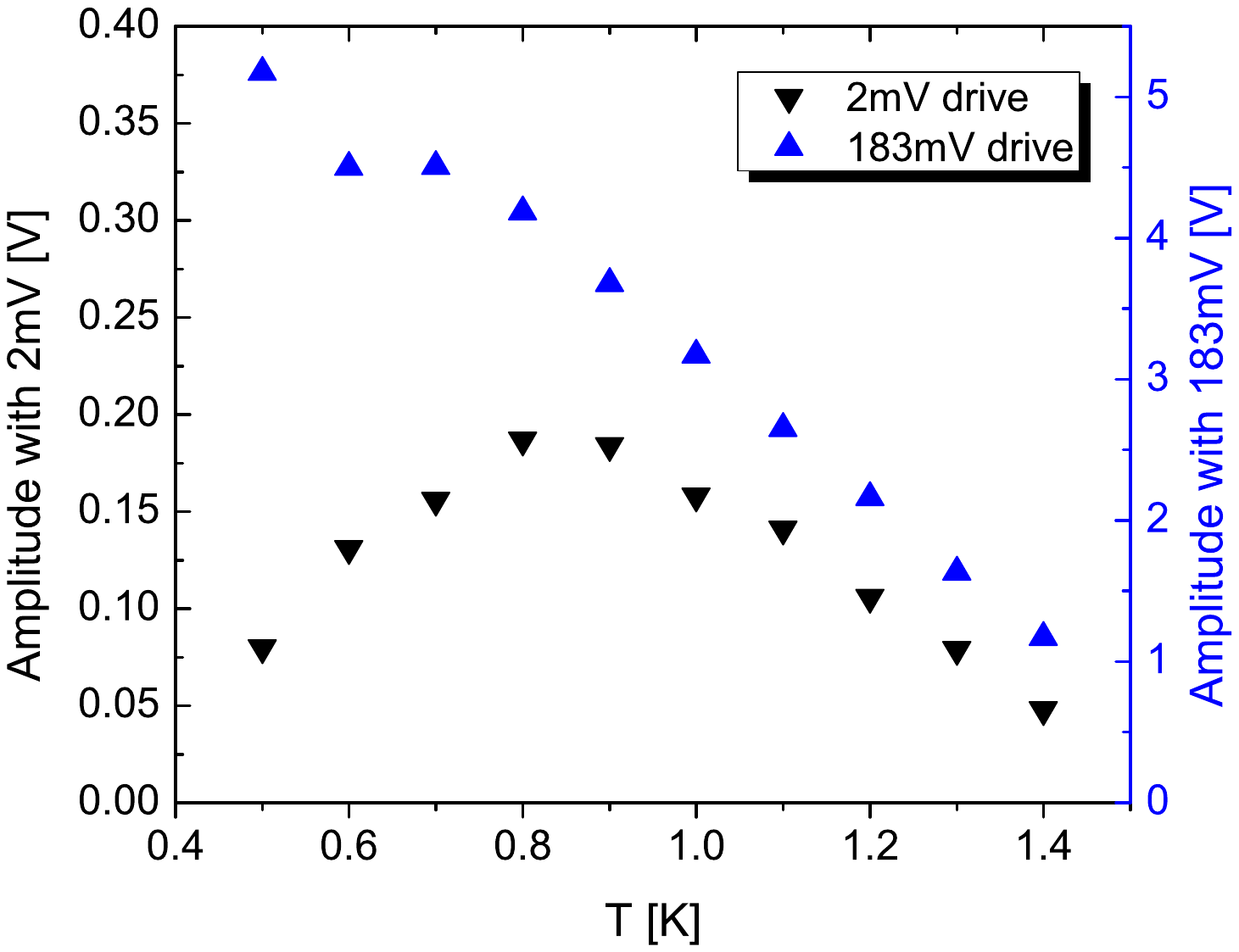} {}
\caption{Temperature dependence of the oscillation amplitude under a constant drive for a $^4$He crystal containing 100ppm $^3$He. Note that the values of the AC voltage driving the TO for the two data sets shown differ by almost 2 orders of magnitude.}
\label{fig:Ampwith3He}
\end{figure}
%\begin{figure}[]
%\includegraphics[width=8in]{amplitude-eps-converted-to.pdf} {}
%\caption{(a) Typical time dependence of the resonant frequency and
%dissipation of a cell containing a bcc crystal during the final
%stages of crystal growth. Initially, the frequency decreases and
%the Q increases as He is added to the cell in small steps. At 7.7
%hours from the beginning of the growth, the cell is full of solid
%and the crystal spontaneously disorders. The resonant frequency
%increases as if 30\% of the He mass has decoupled from the TO. The
%dissipation (Q$^{-1}$) first increases sharply, and then decreases
%with time to about half the maximal value as the crystal reaches a
%steady state. Panel (b) shows the response of the TO during and
%after the disordering stage on an expanded scale.}
%\label{fig:amplitude}
%\end{figure}
%Since the disordered crystal is made up of grains a fraction of a
%mm in size, so as one moves across the 2 mm wide annulus, only a
%few grain boundaries are encountered. So, instead of the continuum
%non-local model of Nussinov et. al. \cite{Nussinov},

An important clue as to the source of dissipation came from studying crystals grown from a $^3$He-$^4$He mixture.
Figure \ref{fig:Ampwith3He} shows the temperature dependence of the resonant amplitude of such a crystal, for two  values of the applied torque. At a small driving torque, the amplitude as a function of temperature shows a peak around T=0.8K.
At high drive, the peak disappears and the resonant amplitude decreases monotonically with temperature. This particular temperature at which the amplitude at low drive shows a maximum has been associated with the onset of binding of $^3$He atoms to dislocations\cite{a025,a006}. Impurity atoms on dislocation cores contribute to the pinning of dislocations, reducing their mobility. It is
therefore likely that the mobility of dislocations below 1K would be
reduced in the presence of $^3$He atoms, as these act as pinning centers. At high
enough drive, dislocations break away from these pinning sites
and the presence of $^3$He does not influence the oscillation amplitude strongly. The detachment of $^3$He from dislocations under high excitation level was observed in other
experiments\cite{a011,a025}. Under these conditions, the presence of $^3$He is felt in other ways which we discuss later on. At any rate, it seems that motion of dislocations strongly influences the dissipation of the TO. On the basis of these observations, we decided to try to analyze the data using a dislocation based dissipation mechanism.

%The binding energy of
%$^3$He atoms to dislocations was estimated as 0.43K
%\cite{KojimaPrb} and 0.7K in another paper\cite{Paalanen}.
To analyze the data, we chose to
use a discrete model, in which the solid helium inside
the annular sample space is divided into 2 parts (see figure
\ref{fig:grains}). One part is in contact with the outer and inner
walls, and contains the solid which moves with the walls. The second
(inner) part contains the solid which is weakly coupled with the
outer band and can therefore move in relation to it. The solid in
the middle band represents the decoupled mass. We assume that all the solid in the inner part is dragged by
friction with the outer part and moves at an angular velocity
$\dot{\theta_{m}}$ proportional to $\dot{\theta}$, the velocity of
the cell. The relative motion between the inner and outer solid bands in our model is represented by a phase
lag $\alpha$, namely $\dot{\theta_{m}}=\beta \dot{\theta} e^{-i
\alpha}$ ($0<\beta<1$ is a constant). The friction force between
the two bands is assumed to depend linearly on
the relative velocity $\dot{\theta}-\dot{\theta_{m}}$, with an
effective friction coefficient
$\gamma_{2}$. Denoting the moment of inertia of the cell including the
solid rigidly attached to it by $I$, we can write the equation of
motion of the cell.
\begin{equation}
I\ddot{\theta}+\gamma_{1} \dot{\theta} +
\gamma_{2}(\dot{\theta}-\beta \dot{\theta} e^{-i \alpha})+ \kappa
\theta = \tau_{0} e^{i \omega t}
\label{eq:I}
\end{equation}
Here $\tau_{0}$ is the external driving torque, $\kappa$ is the
torsion constant of the rod, and $\gamma_{1} \dot{\theta}$ is the
internal friction of the TO and the solid attached to it.

We consider the case where the friction associated with the
intergrain motion is much lower than the internal friction within
the solid moving with the cell, namely $\gamma_{1} \gg \gamma_{2}
(1-\beta cos(\alpha))$.  We also take $\omega_{0} \gg \gamma_{2}
\beta sin(\alpha) / 2 I$. Both these assumption hold for our
system.  In this case, the solution of equation \ref{eq:I} yields the resonant
frequency $\omega_{\alpha}$ and amplitude $\theta_{0}$.
\begin{equation}
\omega_{\alpha} \approx \omega_{0} - \beta \gamma_{2}
sin(\alpha) / 2 I.
\label{eq:omegaa}
\end{equation}
\begin{equation}
\theta_{0} \approx \tau_{0}/ \omega_{0} \gamma_{1} -
\tau_{0}/\omega_{0} \gamma_{1} (\gamma_{2} / \gamma_{1}) (1-\beta
cos(\alpha))
\label{eq:theta0}
\end{equation}

\begin{figure}[]
\includegraphics[width=4in]{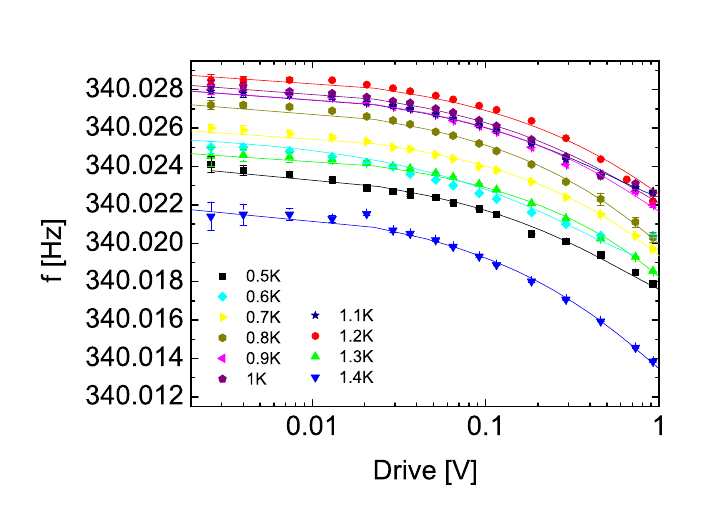} {}
\caption{Resonant frequency of the TO containing an hcp crystal
grown of 100ppm $^3$He-$^4$He mixture, as function of the driving
AC voltage. This particular crystal was initially grown in the bcc
phase at 1.65K and transformed into the hcp structure at 1.46K.
Data sets are for different temperatures.  Solid lines are fits to
equation \ref{eq:omegaa} with $\alpha \propto \sqrt{\tau_{0}}$. }
\label{fig:fvsdrive}
\end{figure}

Here $\omega_{0}=\sqrt{\kappa / I}$. In our model, the phase lag
$\alpha$ should increase with the relative velocity which in turn
increases with the applied torque. To find this dependence, in
figure \ref{fig:fvsdrive} we plot the resonant frequency of the
cell as function of the applied AC voltage. The small offsets
between different curves are due to temperature dependence of
$\omega_{0}$ which are taken as given. The data were fit to
equation \ref{eq:omegaa}. We found that data sets at different
temperatures all have the same dependence on the external torque.
The solid lines shown in figure \ref{fig:fvsdrive} were obtained
by taking the phase lag $\alpha $ to depend linearly on
$\sqrt{\tau_0}$, with no temperature dependence. Obviously, the
fit is good. We remark in this context that in a classical model
of a dislocation gliding with a speed $v$ through a periodic
(Peierls) potential\cite{a026}, the
dislocation radiates energy in the form of sound. This mechanism
causes the speed $v$ to increase as the square root of the applied
stress.

\begin{figure}[]
\includegraphics[width=4in]{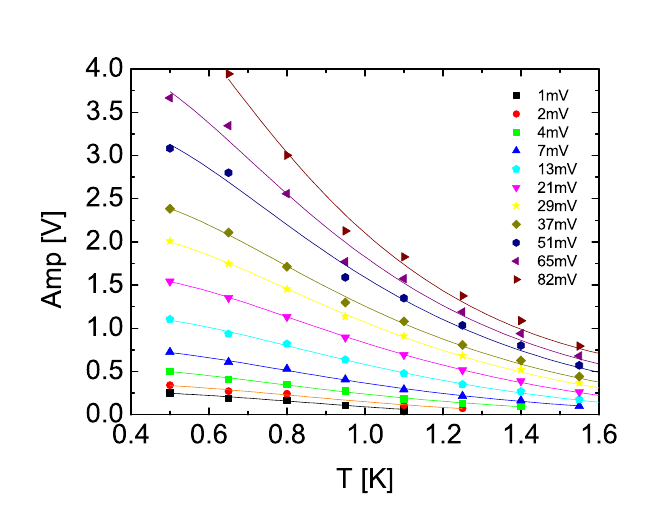} {}
\caption{Resonant amplitude of the TO as function
of temperature, for a $^4$He hcp crystal grown at 1.83K. Each data set is associated with
a constant value of the
drive shown in the legend. Solid lines are fits to Eq. \ref{eq:theta0}, with the
friction coefficient $\gamma_{2}$ taken from Eq.
\ref{eq:gamma2T}.} \label{fig:ampvsT}
\end{figure}

Keeping this in mind, we next fit the data of the resonant amplitude $\theta_{0}$. The
first term in equation \ref{eq:theta0}, $\tau_{0}/ \omega_{0}
\gamma_{1}$, is the amplitude for the case where the solid moves
with the cell. From fitting the data we found that this term is
linear in the external torque and has the same magnitude as the
measured resonant amplitude of the TO with the cell full of an
annealed crystal. Hence, the first term in equation \ref{eq:theta0} represents what would be the
resonant amplitude if the TO included only of the cell and the solid helium attached to
its walls (the outer band). The second term in equation \ref{eq:theta0} decreases the
resonant amplitude due to additional friction between the mobile and immobile solid.  In our model, the
temperature dependence of this term is contained in the
friction coefficient $\gamma_{2}$.
We found that we can fit the temperature
dependence of the oscillation amplitude at a fixed external torque
using $\gamma_{2}$ given by
\begin{equation}
\gamma_{2}=\eta\frac{ e^{-W_{0}/k_{B}T}}{k_{B}T}
\label{eq:gamma2T}
\end{equation}
Here, $\eta$ and $W_{0}$ are fitting parameters. This particular
form describes the rate at which dislocations move during climb via
vacancy emission/absorption by jogs \cite{a027}. The energy $W_0$
is the characteristic energy scale associated with this dissipative process.
In figure \ref{fig:ampvsT} we show the fit of the temperature
dependence of $\theta_{0}$ for a typical set of data, here for an
hcp crystal grown at 1.83K and cooled by 0.15K steps down to 0.5K.
The different sets of data are for different values of the
external torque. Equation \ref{eq:theta0} with $\gamma_{2}$ given
by equation \ref{eq:gamma2T} fits our data very well over the
whole temperature range (0.5K-1.8K), over 2 orders of magnitude in
the driving torque, and for all the crystals which were used in
this work.

\begin{figure}[]
\includegraphics[width=3in]{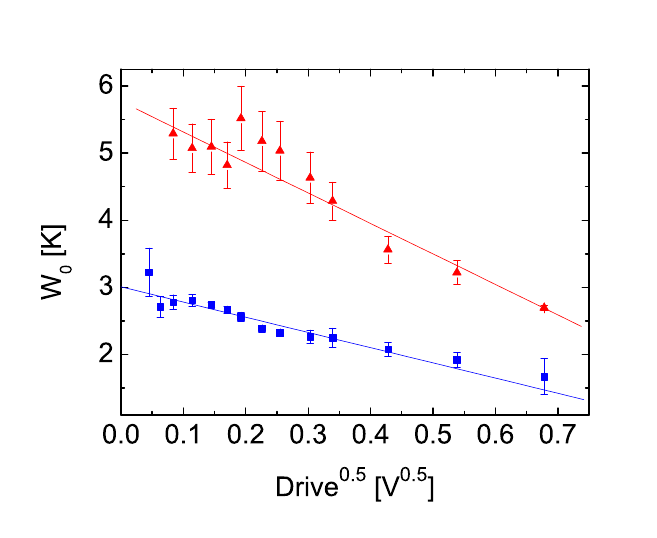} {}
\caption{Characteristic energy $W_{0}$ vs. the square root of the
driving voltage for different crystals. $W_{0}$ was obtained from
fitting the data  to Eq. \ref{eq:gamma2T}. Blue squares are the
combined data of two hcp crystals, one grown at 1.83K and the
second at 1.37K. The red triangles are the combined data of three
crystals, one grown as bcc at 1.69K and then cooled into the hcp
phase through the lower triple point, and two crystals grown from
a 100ppm $^3$He-$^4$He mixture, one at 1.65K and the other at
1.8K. Solid lines are a linear fit. } \label{fig:E0}
\end{figure}

Regarding the energy scale $W_{0}$ which emerges from the fit, Figure \ref{fig:E0} shows the values of $W_{0}$ for all
the crystals in this study. We found that these values fall into two groups. For
hcp crystals grown using commercially pure $^4$He we find
$W_0\sim$3.0K at small stress. For crystals grown initially as bcc
or from a $^3$He-$^4$He mixture, $W_0\sim$5.8K. The fact that
$W_{0}$ takes only two distinct values indicates that crystals
grown by the same procedure grow with a very similar orientation.
By the same token, crystals grown under different conditions can have
somewhat different orientations and a different value of $W_{0}$. In the elastic approximation, the calculated energy of a jog in solid
He is 5.8K \cite{a012}, similar to the values we found. A climb of dislocations
leads to plastic deformation of the solid. The fact that we can fit the amplitude vs. temperature
using a single value of $W_{0}$ suggests that the interface between the moving and static band is the same around the perimeter of the cell. In this case, the crystalline deformation associated with the
stress would be similar over the whole interface between the moving and static solid. This result is consistent with our
direct determination  that
this interface is the basal plane\cite{a016}.
For completeness, we tried to fit the temperature dependence of the amplitude
using other functional forms of $\gamma_{2}$, including among others an Arrhenius type exponential
dependence and a power law in temperature. The extended
temperature range over which we can fit the data enabled us to
clearly discriminate between the various forms, and state that
none of these fit the data nearly as well as that given in Eq.\ref{eq:gamma2T} .

Figure \ref{fig:E0} also shows that $W_{0}$ decreases with stress
in proportion to the square root of the stress (external torque).
Such dependence is predicted within the classical theory of dislocations \cite{a027}, and
results from the dependence of the vacancy formation energy on
the local stress. Climb of dislocations requires an
emission/absorption of vacancies at the core of the dislocation.
The local stress in or near the core is significant, affecting the
formation energy of the vacancy. This dependence of the formation energy was
calculated by Pollet et al.\cite{a028}. In their simulations,
it was found that in solid He under stress the formation energy is
much lower than in a perfect solid, and similar to the values of
$W_{0}$ which we found. Recently, Suhel and Beamish \cite{a029} found an activation
energy of 5.1K characterizing the structural relaxation of an hcp
polycrystal.  All these values, together with the values of $W_{0}$
which we found, set the characteristic energy scale associated
with structural changes in the solid caused by frictional motion.
Since this mechanism of friction involves only elementary crystalline defects, one can perhaps call it the low limit of a slip-stick motion.

In conclusion, it seems that the dissipation due to a relative motion of two solid $^4$He
bodies above 0.5K can be successfully modeled by assuming a
small effective friction at the interface. In disordered single
crystals used in our experiments, this interface is composed of low angle
grain boundaries. The fact that a well defined energy scale $W_{0}$
emerges from fitting the data indicates that the orientation of
all internal interfaces is indeed very similar,  Further, the absence of pinning at these relatively high
temperature allows us to observe response associated with
individual dislocations rather than those of a pinned dislocation
tangle. The low value of the inter-grain dissipation responsible
for the mass decoupling effect is still a puzzle. It can arise
from a superfluid behaviour of the grain boundaries\cite{a030},or
from dislocation glide. Recent measurements\cite{a008} on oriented single crystals show that dislocations glide freely parallel to the basal plane, which is the interface in our experiment\cite{a016}. Such motion can induce a slip of the crystal near the interface, which would appear equivalent to a relative motion of the two solid bands. The dependence of the phase lag angle on the external drive, $\alpha \propto\sqrt{\tau_0}$ which we found
may offer some clue with respect to this question. Regarding the
possibility that most grain boundaries in hcp $^4$He become superfluid at 0.5K as predicted
by Pollet et al.\cite{a030}, we do see that
the dissipation of our TO decreases with decreasing temperature and at 0.5K it equals that of an
empty cell. This means that the friction at 0.5K is zero within our accuracy.
\begin{acknowledgements}
We acknowledge useful discussions with L. A. Melnikovsky.  We thank A. Danzig, O. Scaly, S. Hoida and L. Yumin for their assistance. This work was supported by the Israel Science Foundation and by the Technion Fund for Research.
\end{acknowledgements}
\bibliographystyle{spphys}        % APS-like style for physics
%\bibliography{frictionless_jltp_revised}  % name your BibTeX data base

\end{document}